\documentclass[aps,twocolumn,preprintnumbers,nofootinbib,superscriptaddress]{revtex4-1}

\usepackage{amsmath,amssymb,bm}
\usepackage[dvips]{graphicx}
\usepackage{hyperref}
\usepackage{yfonts}
\usepackage{color}
\newcommand{\beq}{\begin{eqnarray}}
\newcommand{\eeq}{\end{eqnarray}}
\renewcommand\d{\partial}

\begin{document}

\preprint{MIT-CTP/4778}
\title{Nonlinear Chiral Transport Phenomena}
\author{Jiunn-Wei Chen}
\affiliation{Department of Physics, Center for Theoretical Sciences,
and Leung Center for Cosmology and Particle Astrophysics,
National Taiwan University, Taipei 10617, Taiwan}
\affiliation{Center for Theoretical Physics, Massachusetts
Institute of Technology, Cambridge, MA 02139, USA}

\author{Takeaki Ishii}
\thanks{Affiliation until March 2015}
\affiliation{Department of Physics, Keio University, Yokohama 223-8522, Japan}

\author{Shi Pu}
\affiliation{Institute for Theoretical Physics, Goethe University, Max-von-Laue-Str.~1, 60438 
Frankfurt am Main, Germany}

\author{Naoki Yamamoto}
\affiliation{Department of Physics, Keio University, Yokohama 223-8522, Japan}

\begin{abstract}
We study the nonlinear responses of relativistic chiral matter to the external fields, such as 
the electric field ${\bm E}$, gradients of temperature and chemical potential, ${\bm \nabla} T$ 
and ${\bm \nabla} \mu$. Using the kinetic theory with Berry curvature corrections under the 
relaxation time approximation, we compute the transport coefficients of possible new electric 
currents that are forbidden in usual chirally symmetric matter, but are allowed in chirally 
asymmetric matter by parity. In particular, we find a new type of electric current proportional 
to ${\bm \nabla} \mu \times {\bm E}$ due to the interplay between the effects of the Berry 
curvature and collisions. We also derive an analogue of the ``Wiedemann-Franz" law specific 
for anomalous nonlinear transport in relativistic chiral matter.
\end{abstract}
\maketitle

\section{Introduction}
Transport phenomena are abundant in our everyday life and are important in wide areas of 
physics from condensed matter physics and nuclear physics to astrophysics. A familiar 
example is the Ohm's law, ${\bm j}_e = \sigma {\bm E}$, where the electric current flows in 
the direction of the external electric field ${\bm E}$. 
In addition to such first-order transport, various kinds of second-order transport phenomena 
have already been revealed in the 19th century. The well-known examples are the Hall effect 
and Nernst effect, ${\bm j}_e = \sigma_{EB} {\bm E} \times {\bm B}$ and 
${\bm j}_e = \sigma_{TB} (- {\bm \nabla} T) \times {\bm B}$, respectively, where ${\bm B}$ is
the magnetic field and $T$ is temperature. One can question whether other second order transport 
phenomena are possible. For example, one might imagine the electric current of the form 
${\bm j}_e = \sigma_{E \mu} {\bm \nabla} \mu \times {\bm E}$ with $\mu$ the chemical potential. 
However, such a current is not consistent with parity and is usually forbidden in a system that 
respects parity.

In this paper, we argue that such exotic transport phenomena become possible in relativistic matter
with chirality imbalance (which we shall simply refer to as chiral matter below) where parity 
is explicitly violated. Examples of chiral matter are the electroweak plasma in the early 
Universe \cite{Joyce:1997uy}, quark-gluon plasmas created in heavy ion collisions 
\cite{Kharzeev:2007jp, Fukushima:2008xe}, electromagnetic plasmas in neutron stars 
\cite{Charbonneau:2009ax, Ohnishi:2014uea}, neutrino matter in supernovae \cite{Yamamoto:2015gzz}, 
and a new type of materials called the Weyl semimetals \cite{Vishwanath, BurkovBalents, Xu-chern}. 
In this paper, we explicitly compute the transport coefficients of these nonlinear anomalous transport 
in chiral matter at low temperature, based on the kinetic theory with Berry curvature corrections 
\cite{Son:2012wh, Son:2012bg, Stephanov:2012ki, Son:2012zy, Chen:2012ca, Manuel:2014dza} 
under the relaxation time approximation. 

We show that the nonlinear anomalous transport above arises from the interplay between
the Berry curvature and collisions. We also derive a universal relation independent of the relaxation time, 
which is similar to the Wiedemann-Franz law in usual metals, but is specific for nonlinear anomalous 
transport in chiral matter. Our main results are summarized in Eqs.~(\ref{WF}) and (\ref{WF2}).

In this paper, we set $c=k_{\rm B}=1$ unless stated otherwise, but keep $e$ and $\hbar$ explicitly.

\section{Classification of currents from symmetries}
We first classify the possible electric currents in the presence of various external 
fields to the second order in derivatives. Although we limit ourselves to the electric 
currents in this paper, the same classification is also applicable to heat currents.
The external fields we consider here are the electric field ${\bm E}$, 
magnetic field ${\bm B}$, gradients of temperature and chemical potential 
${\bm \nabla} T$ and ${\bm \nabla} \mu$, and their possible combinations.
(In this paper, we do not consider external fields involving the fluid velocity 
${\bm v}$, such as the vorticity ${\bm \omega} = {\bm \nabla} \times {\bm v}$.)
We also assume external fields are time-independent.
We will see that the ${\cal P}$ (parity), ${\cal C}$ (charge conjugation), and 
${\cal T}$ (time reversal) symmetries put stringent constraints on the possible 
transport phenomena.

Let us first consider the electric currents in usual parity-invariant systems.
To the second order derivatives, the general expression that is consistent with 
${\cal CPT}$ symmetries reads ${\bm j}_{+} = {\bm j}_+^{(1)} + {\bm j}_+^{(2)}$,
where
\begin{align}
\label{j+1}
{\bm j}_+^{(1)} &= \sigma_E {\bm E} + \sigma_T (- {\bm \nabla} T)
+ \sigma_{\mu} (- {\bm \nabla} \mu), \\
\label{j+2}
{\bm j}_+^{(2)} &= \sigma_{EB} {\bm E} \times {\bm B}
+ \sigma_{TB} (- {\bm \nabla} T) \times {\bm B} \nonumber \\
& \qquad + \sigma_{\mu B} (- {\bm \nabla} \mu) \times {\bm B}.
\end{align}
Here and henceforth the subscript $\pm$ denotes the ${\cal P}$-even and 
${\cal P}$-odd transport coefficients. The upper indices denote the number 
of derivatives. 

Note that the transport coefficients in Eq.~(\ref{j+1}) are ${\cal T}$-odd and must be
accompanied with dissipation: under ${\cal T}$, ${\bm j} \rightarrow -{\bm j}$, 
${\bm E} \rightarrow {\bm E}$, and ${\bm B} \rightarrow -{\bm B}$.
On the other hand, the transport coefficients in Eq.~(\ref{j+2}) are ${\cal T}$-even 
and do not necessarily involve dissipation (see also Ref.~\cite{Kharzeev:2011ds}). 
For the electric current, the $\sigma_E$ term is called the Ohm's law, 
the $\sigma_T$ term the Seebeck effect, the $\sigma_{EB}$ term the Hall effect, 
and the $\sigma_{TB}$ term the Nernst effect. 

Let us turn to a parity-violating system with nonzero chiral chemical potential, 
$\mu_5 = (\mu_{\rm R} -\mu_{\rm L})/2$. Under ${\cal P}$, the chiral chemical potential 
transforms as  $\mu_5 \rightarrow -\mu_5$.
Then one can write down the general expression for the parity-odd current
to the second order. For simplicity, we first consider the system only with 
right-handed fermions at finite chemical potential $\mu_{\rm R}$ 
(which we shall denote $\mu$ in the following). Then the current is expressed by 
${\bm j}_{-} = {\bm j}_-^{(1)} + {\bm j}_-^{(2)}$, where
\begin{align}
\label{j-1}
{\bm j}_-^{(1)} &= \sigma_B {\bm B},
\\
\label{j-2}
{\bm j}_-^{(2)} &= \sigma_{ET} {\bm E} \times (- {\bm \nabla} T) 
+ \sigma_{E \mu} {\bm E} \times (- {\bm \nabla} \mu)  \nonumber \\
& \qquad + \sigma_{T \mu} (- {\bm \nabla} T) \times (- {\bm \nabla} \mu),
\end{align}
where all the transport coefficients in Eqs.~(\ref{j-1}) and (\ref{j-2}) are functions of $\mu$.
The $\sigma_B$ term is called the chiral magnetic effect 
\cite{Vilenkin:1980fu, Nielsen:1983rb, Alekseev:1998ds, Son:2004tq, Fukushima:2008xe}.
Note that the transport coefficient $\sigma_B$ (which is referred to as the chiral magnetic 
conductivity) in Eq.~(\ref{j-1}) is ${\cal T}$-even and is indeed dissipationless (see below) 
while those in Eq.~(\ref{j-2}) are ${\cal T}$-odd and are dissipative. 

It has been revealed that the coefficient $\sigma_B$ is uniquely fixed from the constraint 
of the second law of thermodynamics and that it is related to the coefficient of the chiral 
anomaly \cite{Son:2009tf}; its relation to the chiral anomaly also underlies that this current 
is dissipationless.
However, it is a nontrivial question whether other parity-violating terms in Eq.~(\ref{j-2}) 
are fixed uniquely by the anomaly coefficients. We shall show that these new transport 
phenomena arise due to an interplay between the topological terms (the Berry curvature) 
and collisional terms.

\section{Kinetic theory}
We will be interested in the nonlinear electric currents that arise due to the explicit 
violation of parity symmetry in the system, shown in Eq.~(\ref{j-2}). As there
is no such a current involving the magnetic field ${\bm B}$ there, it will be sufficient 
to consider the case only with ${\bm E}$, ${\bm \nabla} T$, and ${\bm \nabla} \mu$, 
but without ${\bm B}$ for this purpose.

\subsection{Kinetic theory with Berry curvature}
We first briefly review the kinetic theory for a single chiral fermion at $\mu \gg T$
\cite{Son:2012wh, Son:2012bg, Stephanov:2012ki, Son:2012zy, Chen:2012ca, Manuel:2014dza}.
(We will consider a system with both right and left-handed fermions later.) The chiral 
fermions near the Fermi surface possess a Berry curvature in momentum space 
\cite{Volovik, Son:2012wh}. The equations of motion for chiral quasiparticles in an 
electric field ${\bm E}$ and the Berry curvature ${\bm \Omega}_{\bm p}$ are \cite{SundaramNiu}
\begin{align}
\dot {\bm x} &= {\bm v} + \dot {\bm p} \times {\bm \Omega}_{\bm p},
\\ 
\dot {\bm p} &= e{\bm E},
\end{align}
where ${\bm v} = \d \epsilon_{\bm p}/\d {\bm p}$. Substituting them into 
the Boltzmann equation for the distribution function $n_{\bm p}({\bm x})$,
\beq
\frac{\d n_{\bm p}}{\d t} + \dot {\bm x} \cdot \frac{\d n_{\bm p}}{\d {\bm x}}
+ \dot {\bm p} \cdot \frac{\d n_{\bm p}}{\d {\bm p}} = I_{\rm coll} \{ n_{\bm p} \}\,,
\eeq
the kinetic equation in the present case is given by
\cite{Son:2012wh, Son:2012bg, Stephanov:2012ki, Son:2012zy, Chen:2012ca, Manuel:2014dza} 
\beq
\label{kinetic}
\frac{\d n_{\bm p}}{\d t} + ({\bm v} + e {\bm E} \times {\bm \Omega}_{\bm p}) 
\cdot \frac{\d n_{\bm p}}{\d {\bm r}} 
+ e {\bm E} \cdot \frac{\d n_{\bm p}}{\d {\bm p}} = I_{\rm coll} \{ n_{\bm p} \}\,, \nonumber \\
\eeq
where $I_{\rm coll} \{ n_{\bm p} \}$ is the collision term.

\subsection{Relaxation time approximation}
For the collision term in Eq.~(\ref{kinetic}), we use the relaxation time approximation,
\beq
\label{RTA}
I_{\rm coll} = -\frac{\delta n_{\bm p}}{\tau},
\eeq
where $\tau$ is the relaxation time (which we assume to be a constant) and 
$\delta n_{\bm p} = n_{\bm p} - n_{\bm p}^{(0)}$ is the deviation from the equilibrium 
distribution function.

For inhomogeneous temperature $T$ and chemical potential $\mu$, 
the stationary solution of the kinetic equation can be found 
order by order in derivatives,
\begin{subequations}
\label{dn}
\begin{align}
n_{\bm p}^{(0)} &= \frac{1}{e^{(\epsilon - \mu)/T}+1}\,,
\\
\delta n_{\bm p}^{(1)} &= \tau {\bm v} \cdot 
\left(- e {\bm E} + {\bm \nabla}\mu + \frac{\epsilon - \mu}{T} {\bm \nabla}T \right)
\frac{\d n_{\bm p}^0}{\d \epsilon}\,,
\\
\delta n_{\bm p}^{(2)} &= \tau e {\bm E} \times {\bm \Omega}_{\bm p}
\cdot \left({\bm \nabla}\mu + \frac{\epsilon - \mu}{T} {\bm \nabla}T \right)
\frac{\d n_{\bm p}^0}{\d \epsilon}\,.
\end{align}
\end{subequations}
Here $n_{\bm p}^{(0)}$ is the equilibrium Fermi-Dirac distribution
and the upper indices $k=0,1,2$ denote the terms involving 
$k$ derivatives. Precisely speaking, there are other terms not listed in
$\delta n_{\bm p}^{(2)}$ which are not related to the Berry curvature corrections.
However, they are irrelevant for our purpose of computing the parity-violating
transport coefficient to the second order and we will ignore them below.

We pause here to remark on the relaxation time approximation in Eq.~(\ref{RTA}). 
Performing the momentum integral of the kinetic equation (\ref{kinetic}), one 
obtains the continuity equation as long as the following condition is fulfilled:
\beq
\int \frac{d{\bm p}}{(2\pi \hbar)^3} I_{\rm coll} \{ n_{\bm p} \} = 0.
\eeq
One can check that the collision term in Eq.~(\ref{RTA}) together with Eq.~(\ref{dn}) indeed 
satisfies this condition, at least for a spherically symmetric Fermi sphere; so the relaxation 
time approximation in the present case is consistent with the particle number conservation law.

\section{Nonlinear electric currents}

\subsection{System with a single chiral fermion}
To simplify the argument, we first consider the case with a single chiral fermion.
The electric current in the presence of the Berry curvature corrections is given by 
\cite{Son:2012wh, Son:2012zy}
\beq
\label{j}
{\bm j}_e = e \int \frac{d{\bm p}}{(2\pi \hbar)^3} 
\left[{\bm v} n_{\bm p} + (e{\bm E} \times {\bm \Omega}_{\bm p}) n_{\bm p} -
\epsilon_{\bm p} {\bm \Omega}_{\bm p} \times \frac{\d n_{\bm p}}{\d {\bm x}}
\right], \nonumber \\
\eeq
The first term in Eq.~(\ref{j}) is the usual convective current,
the second term is the anomalous Hall current, and the third term is the 
magnetization current \cite{Chen:2014cla} that originates from the magnetic 
moment of chiral fermions \cite{Son:2012zy, Manuel:2014dza}.

The electric current to the second order in derivatives is then
\begin{align}
{\bm j}_e^{(2)} \! &= e \! \int \! \! \frac{d{\bm p}}{(2\pi \hbar)^3} \! \!
\left[{\bm v} \delta n_{\bm p}^{(2)} \! + \! (e{\bm E} \times {\bm \Omega}_{\bm p}) \delta n_{\bm p}^{(1)} \! - \!
\epsilon_{\bm p} {\bm \Omega}_{\bm p} \! \times \! \frac{\d \delta n_{\bm p}^{(1)}}{\d {\bm x}}
\right] \nonumber \\
& \equiv {\bm j}_1 + {\bm j}_2 + {\bm j}_3.
\end{align}
They can be respectively computed by substituting Eq.~(\ref{dn}).
For right-handed fermions, for example, the results read 
\begin{subequations}
\begin{align}
{\bm j}_1
& = -\frac{e^2 \tau}{12 \pi^2 \hbar^2} {\bm \nabla} \mu \times {\bm E}, 
\\
{\bm j}_2
& = \frac{e^2 \tau}{12 \pi^2 \hbar^2} {\bm \nabla} \mu \times {\bm E}, 
\\
{\bm j}_3 
& = \frac{e^2 \tau}{12 \pi^2 \hbar^2} {\bm \nabla} \mu \times {\bm E}.
\end{align}
\end{subequations}
Summing over the three contributions above, we obtain
\beq
\label{sigma_Emu}
\sigma_{E \mu} = \pm \frac{e^2 \tau}{12 \pi^2 \hbar^2}\,,
\eeq
for right and left-handed fermions, respectively.

Note here that two contributions of the form ${\bm \nabla} T \times {\bm \nabla} \mu$
with the opposite signs exactly cancel out in ${\bm j}_3$, and the electric current 
proportional to ${\bm \nabla} T \times {\bm \nabla} \mu$ is absent, at least within the
present relaxation time approximation (although it is in principle allowed by the symmetry).
We also find that the electric current proportional to ${\bm \nabla} T \times {\bm E}$ 
disappears after the momentum integral.%
\footnote{Precisely speaking, the result of the momentum integral can be different in Weyl 
metals in condensed matte systems, where the description of chiral fermions has finite UV 
and IR energy cutoffs. In this case, $\sigma_{TE}$ can remain nonzero at finite temperature, 
but is suppressed exponentially as $\sigma_{ET} \propto e^{-\mu/T}$. In particular, 
$\sigma_{ET}$ vanishes at $T = 0$.}
At this moment, we do not have a clear understanding of the physical reason that 
underlies their absence. 

As a comparison, let us also compute the electrical conductivity in relativistic matter.
Under the relaxation time approximation, we obtain the Ohmic current from 
Eqs.~(\ref{dn}) and (\ref{j}) as
\beq
{\bm j}_e^{(1)} = -e^2 \tau \int \frac{d{\bm p}}{(2\pi \hbar)^3} {\bm v} ({\bm v}\cdot {\bm E}) \frac{\d n_{\bm p}^0}{\d \epsilon}
= \frac{e^2 \mu^2 \tau}{6 \pi^2 \hbar^3} {\bm E}\,.
\eeq
Using the number density for a single chiral fermion, $n = \mu^3/(6\pi^2 \hbar^3)$, 
the electrical conductivity can be expressed as
\beq
\label{sigma_E}
\sigma_E = \frac{n e^2 \tau}{\mu}\,.
\eeq
This takes the familiar form of the Drude-type formula if we identify the effective mass 
$m^*=\mu$.

By taking the ratio between Eqs.~(\ref{sigma_Emu}) and (\ref{sigma_E}), 
we obtain the analogue of the ``Wiedemann-Franz law" specific for relativistic chiral matter:
\beq
\label{WF}
\frac{\mu^2  |\sigma_{E \mu}|}{\sigma_E} = \frac{\hbar c}{2}\,,
\eeq
where we restored $c$.
Equation (\ref{WF}) shows that the ratio between the electrical conductivity $\sigma_E$ and 
anomalous nonlinear conductivity $\sigma_{E\mu}$ (multiplied by $\mu^2$ to match the dimension) 
is a universal quantity that depends only on the physical constants $\hbar$ and $c$. At least within 
the present relaxation time approximation, this relation is independent of the microscopic details 
(i.e., the relaxation time $\tau$). 

It should be remarked that, in the case of usual metals, the Wiedemann-Franz-type law 
for the ratio between the linear and nonlinear transport coefficients of electric currents do 
not exist. This may be understood as follows: since all the nonlinear transport $\sigma_{EB}$, 
$\sigma_{TB}$, and $\sigma_{\mu B}$ in Eq.~(\ref{j+2}) are ${\cal T}$-even, the transport 
coefficients are even functions of $\tau$. Hence, the ratios between these nonlinear transport 
coefficients and the electrical conductivity $\sigma_E \propto \tau$ must depend on $\tau$ 
(i.e., nonuniversal). In chiral matter, on the other hand, the nonlinear transport 
$\sigma_{E \mu}$ is ${\cal T}$-odd and $\sigma_{E \mu}/\sigma_E$ can be independent of $\tau$.

\subsection{System with both right and left-handed fermions}
So far we have considered the case with a single chiral fermion.
We now consider a system with right and left-handed fermions in the presence of 
finite chiral chemical potential $\mu_5 = (\mu_{\rm R} - \mu_{\rm L})/2$. 
We denote the vector chemical potential by $\mu = (\mu_{\rm R} + \mu_{\rm L})/2$.
In this case, the kinetic equation (\ref{kinetic}) needs to be extended to include both 
right and left-handed fermions as (see also Refs.~\cite{Son:2012bg, Stephanov:2014dma})
\beq
\frac{\d n_{\bm p}^{i}}{\d t} \! +  \! ({\bm v} + e {\bm E} \times {\bm \Omega}_{\bm p}) \!
\cdot \! \frac{\d n_{\bm p}^{i}}{\d {\bm r}} 
+ e {\bm E} \! \cdot \! \frac{\d n_{\bm p}^{i}}{\d {\bm p}} \! = \! I_{\rm coll}^{i} \{ n_{\bm p}^{i} \} \,.
\nonumber \\
\eeq
where $i = {\rm R, L}$ denote the chirality of fermions. The collision term $I_{\rm coll}^{i} \{ n_{\bm p}^{i} \}$ 
generally describes the inter-chiral and intra-chiral scatterings. Here we assume that 
the former mean free time (which we denote $\tau_{\rm tr}$) is much larger than the latter 
(which we denote $\tau$) \cite{Son:2012bg, Stephanov:2014dma} and we ignore the effects 
of the former in the leading order in $\tau/\tau_{\rm tr} \ll 1$. 
Then the kinetic equations for right and left-handed fermions are decoupled from each other. 
Analogously to the discussion above, we use the relaxation time approximation,
\beq
I_{\rm coll}^{i} = -\frac{\delta n_{\bm p}^{i}}{\tau}\,,
\eeq
where the thermal relaxation time $\tau$ is assumed to be the same constant for right and 
left-handed fermions \cite{Son:2012bg, Stephanov:2014dma}.%
\footnote{The assumption that $\tau$ is the same for right and left-handed fermions should 
be approximately valid, at least in the regime $\mu_5 \ll \mu$.}
Taking the summation and subtraction of the contributions from right and left-handed fermions, 
we then obtain the nonlinear anomalous electric and axial currents,
\begin{align}
\label{sigma_Emu5}
{\bm j}_e & = {\bm j}_e^{\rm R} + {\bm j}_e^{\rm L} 
= \frac{e^2 \tau}{6 \pi^2 \hbar^2} {\bm \nabla} \mu_5 \times {\bm E}\,, 
\\
{\bm j}_5 & = {\bm j}_e^{\rm R} - {\bm j}_e^{\rm L} 
= \frac{e^2 \tau}{6 \pi^2 \hbar^2} {\bm \nabla} \mu \times {\bm E}\,,  
\end{align}
respectively. 
On the other hand, one finds that the Ohmic current in this case is 
\beq
\label{sigma_EV}
{\bm j}_e = \frac{e^2 \tau (\mu^2 + \mu_5^2)}{3 \pi^2 \hbar^3} {\bm E}\,.
\eeq
Denoting the transport coefficients in Eqs.~(\ref{sigma_Emu5}) and (\ref{sigma_EV}) by 
$\sigma_{E \mu_5}$ and $\sigma_{E}$, respectively, we arrive at the 
Wiedemann-Franz-type law in this case as
\beq
\label{WF2}
(\mu^2 + \mu_5^2) \frac{\sigma_{E \mu_5}}{\sigma_E} = \frac{\hbar c}{2} \,,
\eeq
where we restored $c$ again.


\section{Conclusions}
In this paper, we explored nonlinear responses of chiral matter to external fields, 
based on the kinetic theory with Berry curvature corrections. 
The exotic transport phenomena found in this paper should be relevant to the dynamical 
evolution of chiral matter, such as the electroweak plasma in the early Universe, 
quark-gluon plasmas created in heavy ion collisions, and supernova explosions. 
Our predictions may also be tested experimentally in Weyl semimatals.

In this paper, we derived the analogue of the ``Wiedemann-Franz" law for anomalous
transport, as shown in Eqs.~(\ref{WF}) and (\ref{WF2}). To what extent this relation is 
universal (i.e., independent of microscopic details of systems) beyond the relaxation time 
approximation would be an important question to be investigated in future.

It would be interesting to study possible new nonlinear heat currents specific for 
chiral matter, similar to the nonlinear anomalous electric currents found in this paper. 
One should be able to compute such heat currents by $j_{\rm Q}^i = T^{0i} - \mu j_n^i$, 
where $j_n^i$ is the particle number current and $T^{\mu \nu}$ is the energy-momentum 
tensor including the Berry curvature corrections defined in Ref.~\cite{Son:2012zy}. 
One can also ask the possible effects of finite fermion mass (see, e.g., Ref.~\cite{Chen:2013iga}). 
We defer these questions to future work.

\acknowledgments
J.-W.~C would like to thank the Rudolph Peierls Centre for Theoretical Physics of the 
University of Oxford and Oxford Holography group for hospitality.
J.-W.~C. is supported in part by the Ministry of Science and Technology, Taiwan under 
Grants Nos.~102-2112-M-002-013-MY3 and 105-2918-I-002 -003, and the CASTS of NTU. 
The author S.~P. is supported by the Alexander von Humboldt Foundation, Germany.
The work of N.~Y. is supported in part by JSPS KAKENHI Grants No.~26887032 and 
MEXT-Supported Program for the Strategic Research Foundation at Private Universities, 
``Topological Science'' (Grant No.~S1511006). 

{\it Note added.}---During this work was being completed, we learned that I.~Shovkovy and his 
collaborators also obtained the results \cite{Gorbar2016} that have some overlap with our calculations.

\end{document}